\documentclass[twocolumn]{aastex7}

\begin{document}

\title{Carbon-rich Sub-Neptune Interiors Are Compatible with JWST Observations}

\correspondingauthor{Zifan Lin}
\email{zifanlin@mit.edu}

\author[0000-0003-0525-9647, gname=Zifan, sname=Lin]{Zifan Lin}
\affiliation{Department of Earth, Atmospheric, and Planetary Sciences, Massachusetts Institute of Technology, 77 Massachusetts Avenue, Cambridge, MA 02139, USA}
\email{zifanlin@mit.edu} 

\author[0000-0002-6892-6948, gname=Sara, sname=Seager]{Sara Seager}
\affiliation{Department of Earth, Atmospheric, and Planetary Sciences, Massachusetts Institute of Technology, 77 Massachusetts Avenue, Cambridge, MA 02139, USA}
\affiliation{Department of Physics and Kavli Institute for Astrophysics and Space Research, Massachusetts Institute of Technology, Cambridge, MA 02139, USA}
\affiliation{Department of Aeronautics and Astronautics, MIT, 77 Massachusetts Avenue, Cambridge, MA 02139, USA}
\email{seager@mit.edu}


\begin{abstract}

Many possible interior compositions exist for sub-Neptunes:
ice-poor, ice-rich, and water-dominated interiors can all match the measured masses and radii. Motivated by recent theory of carbon-rich planet formation outside of the refractory organic carbon ``soot line'' and observations of carbon-rich protoplanetary disks around late M dwarfs, we propose another possible sub-Neptune composition: a carbon-rich composition consisting of an iron-silicate core, a carbon layer, and a hydrogen/helium-dominated envelope.
We show that the interiors of three prototypical sub-Neptunes with high-quality spectral observations -- TOI-270 d, GJ 1214 b, and K2-18 b -- are consistent with carbon-rich compositions if they have $\leq100\times$ solar metallicity atmospheres. We further show that carbon-rich interiors lead to atmospheric compositions that match HST and JWST observations. Simulated carbon-rich TOI-270 d transmission spectra pass the $\chi^2$ test under a wide range of C/O, haze, and cloud scenarios. K2-18 b spectral models are broadly consistent with observation, but requires additional sources for carbon species to be fully compatible. GJ 1214 b models, however, are incompatible with observations, ruling out a carbon-rich interior composition, if the atmosphere of the planet is primordial and reflects interior C/O.

\end{abstract}



\section{Introduction} \label{section:intro_carbon}
The interior compositions of sub-Neptunes are highly uncertain. Typical sub-Neptune interior models involve some combination of an iron-silicate core, an ice mantle, and a hydrogen and helium (H/He) envelope. Due to their intermediate bulk densities, vastly different interior compositions can all fit the mass-radius (M-R) measurements of sub-Neptunes, including ice-poor, ice-rich, and ``water world'' compositions
\citep[see e.g.,][]{rogers_three_2010, madhusudhan_interior_2020, rigby_ocean_2024, schmidt_comprehensive_2025}.

Here, we consider another possible sub-Neptune interior composition: carbon-rich composition, where there is a layer of carbon between the iron-silicate core and the H/He-dominated envelope. \cite{kuchner_extrasolar_2005} proposed that ``carbon planets'' can form in protoplanetary disks with high C/O ratios ($>0.98$; for comparison, the solar C/O ratio is 0.55), where carbon and carbides would condense before silicates.
The possibility that 55 Cancri e is a carbon planet \citep{madhusudhan_possible_2012} and the implications of carbon-rich interior on habitability \citep{unterborn_role_2014} were explored, motivated by measurements of high C/O ratios ($>1$) of exoplanet host stars \citep[e.g.,][]{bond_compositional_2010, delgado_mena_chemical_2010}. However, more up-to-date measurements found that C/O ratios of FGK stars in the solar neighborhood are homogeneous and $<0.8$, challenging earlier claims of high C/O stars \citep[e.g.,][]{suarez-andres_co_2018, bedell_chemical_2018}.

Nevertheless, it is still possible to form carbon-rich planets
in disks with Sun-like C/O ratios. Sequential condensation models predicted the formation of carbon-rich planetesimals in disks with low C/O ratios $\sim0.65$ \citep{moriarty_chemistry_2014}, much lower than previous models assuming static equilibrium chemistry. \cite{bergin_exoplanet_2023} posited a ``soot line''
outside of which carbon-rich materials are not destroyed but are available for planet formation \citep[see also][]{kress_soot_2010, li_earths_2021}. Planets formed outside the soot line but within the water ice line would be rich in carbon, depleted in water, and have high C/O ratios even if the star's C/O ratio is Sun-like. \cite{bergin_exoplanet_2023} adopted a conservative estimate for carbon content (0.1--1.0 wt\%, assuming substantial volatile loss during and after formation). The maximum amount of carbon-rich soot assuming efficient volatile retention, however, can be as high as 12--24 wt\%.

Carbon-rich stellar compositions have not been ruled out for M dwarfs, which are not well-constrained chemically due to observational challenges
\citep[e.g.,][]{jahandar_chemical_2024, melo_stellar_2024}. Spitzer revealed a possible link between cool stars and high carbon content in disks. Brown dwarf disks were found to have higher C$_2$H$_2$/HCN and HCN/H$_2$O flux ratios than T Tauri disks, implying a higher C/O ratio in brown dwarf disks \citep{pascucci_atomic_2013}. More recently, JWST observations of protoplanetary disks around very low-mass young M stars revealed carbon-rich chemistry and high gas C/O ratios \citep{tabone_rich_2023, arabhavi_abundant_2024, kanwar_minds_2024}. \cite{long_first_2025} observed a 30-Myr-old disk around a M4.5 star and detected a spectrum dominated by hydrocarbons, suggesting carbon-rich conditions (gas-phase C/O$\gtrsim2$) can persist to a late stage in disk evolution. M dwarfs are the host stars of many well-characterized sub-Neptunes, including the three prototypical sub-Neptunes we highlight in this study: TOI-270 d (host TOI-270 is an M3.0 star), GJ 1214 b (host GJ 1214 is an M4.5 star), and K2-18 b (host K2-18 is an M2.5 star). Note that, however, the disk observations discussed above focus on late ($>$M4) M stars, and only GJ 1214 falls into this category. Nevertheless, these disk observations motivate the exploration of carbon-rich compositions for sub-Neptunes orbiting M dwarfs.


In this paper, we test whether carbon-rich interior compositions can reproduce observations, with a combination of interior, atmosphere, and spectra models. In Section \ref{section:methods_carbon}, we describe our modeling methods. In Section \ref{section:results_carbon}, we summarize our interior and atmosphere modeling results. Implications of our results and a summary of our conclusions are presented in Section \ref{section:discussion_carbon}.

\section{Methods} \label{section:methods_carbon}
Here, we introduce our planetary interior model (Section \ref{section:method_carbon_interior}), atmospheric chemistry model (Section \ref{section:method_carbon_chem}), and transmission spectra model (Section \ref{section:method_carbon_spectra}).

\subsection{Planetary Interior Model} \label{section:method_carbon_interior}
To model carbon-rich sub-Neptune interiors, we use a newly developed planetary interior code package \texttt{CORGI} \citep{lin_interior_2025}. \texttt{CORGI} solves the interior structure of a nonrotating, spherically symmetric planet using three fundamental equations \citep{zapolsky_mass-radius_1969}, namely the mass of a spherical shell
\begin{equation} \label{eq:mass_in_shell}
    \frac{dm(r)}{dr} = 4 \pi r^2 \rho(r),
\end{equation}
hydrostatic equilibrium
\begin{equation} \label{eq:hydro_eq}
    \frac{dP(r)}{dr} = -\frac{Gm(r)\rho(r)}{r^2},
\end{equation}
and equation of state (EOS)
\begin{equation} \label{eq:eos}
    P(r) = f(\rho(r), T(r)).
\end{equation}
The EOS further depends on a temperature profile $T(r)$, or equivalently $T(P)$. We assume an adiabatic pressure-temperature (P-T) profile for the envelope. The iron, silicate, and carbon layers are assumed to be isothermal because the densities of solids are insensitive to the change in temperature \citep{seager_massradius_2007}.

We have made two updates to the original \texttt{CORGI} model described in \cite{lin_interior_2025} to fit the need of this work. The first major update is the inclusion of carbon EOS. At low pressures, we adopt the third-order Birch-Murnagham EOS for graphite from \cite{seager_massradius_2007}. At high pressures, we employ the diamond EOS from \cite{swift_equation_2022}, which was calculated using density functional theory and agrees well with experimental data. We follow \cite{madhusudhan_possible_2012} to assume that graphite-diamond phase transition occurs at a pressure of 10 GPa. At extreme pressures ($\sim1$ TPa), diamond is theoretically predicted to transform into a BC8 structure, but experiments revealed that diamond remains stable up to 2 TPa \citep{lazicki_metastability_2021}. Therefore, we ignore the phase transition of diamond above $\sim1$ TPa pressure, which is well beyond the typical pressures in sub-Neptune carbon layers. The carbon EOS is validated in Appendix \ref{sec:app_model_validation}.

The second update is that we employ a fourth-order Runge-Kutta method to numerically solve for the differential equations of planetary interiors \citep[see][]{rogers_framework_2010}. We assume that carbon-rich sub-Neptunes have a four-layer structure: an iron layer, a silicate layer, a carbon layer, and an H/He dominated envelope with solar metallicity (the default \texttt{CORGI} EOSs are employed for these layers, see \citealt{lin_interior_2025}). We further assume that the iron mass fraction in the iron-silicate core, $x_{\rm Fe}/(x_{\rm Fe}+x_{\rm silicate})$, is fixed at an Sun-like value of 0.332 \citep{adibekyan_compositional_2021}. This assumption is justified by the elemental composition homogeneity in the solar neighborhood \citep{bedell_chemical_2018}, with the exception of a category of high-density planets that have much higher Fe/Si than the Galactic mean \citep{2024Cambioni_inprep, lin_most_2025}. Thus, there are only two compositional free variables, $x_{\rm core} = x_{\rm Fe} + x_{\rm silicate}$ and $x_{\rm carbon}$ (the mass fraction of the envelope $x_{\rm env} = 1 - x_{\rm core} - x_{\rm carbon}$ and hence is not a free variable), while there are two outer boundary conditions $M_p$ and $R_p$. This system is therefore determined, allowing us to use the Runge-Kutta method to solve the differential equations of planetary interiors for probable interior conditions instead of the Markov chain Monte Carlo method employed in \cite{lin_interior_2025}.

\begin{figure*}[ht!]
\centering
\includegraphics[width=\textwidth]{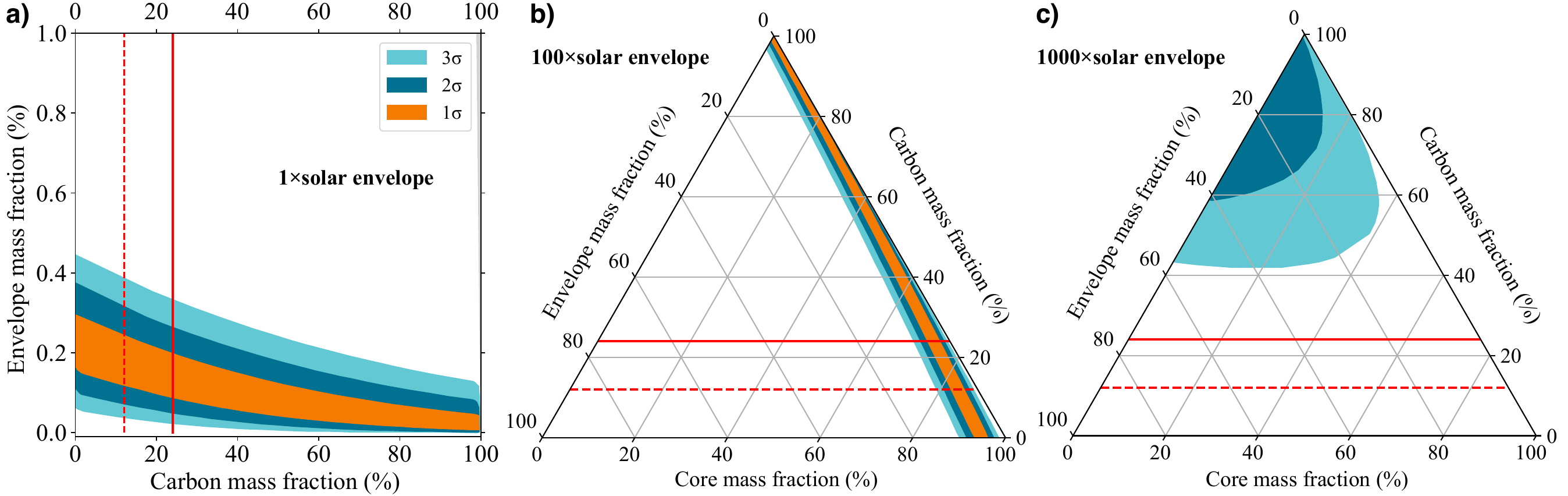}
\caption{Possible interior compositions of TOI-270 d. Compositions consistent within the $1\sigma$, $2\sigma$, and $3\sigma$ M-R uncertainties are shown in orange, dark teal, and light teal, respectively. The red dashed and solid lines show $x_{\rm carbon}=12$ wt\% and 24 wt\%, respectively, the upper bounds of $x_{\rm carbon}$ assuming no volatile loss \citep{bergin_exoplanet_2023}. $x_{\rm carbon}$ defines the bulk carbon content of the planet and in turn defines the C/O ratio in the atmosphere, under the assumption that the atmospheric C/O ratio is the same as the bulk solid C/O ratio. (a) Carbon-envelope diagram for $1\times$ solar metallicity envelope. 
(b) Ternary diagram for $100\times$ solar metallicity envelope. (c) Ternary diagram for $1000\times$ solar metallicity envelope. The $100\times$ solar metallicity scenario allows the widest range of interior compositions and is broadly consistent with the metallicity retrieved by \cite{benneke_jwst_2024}.}
\label{fig:toi270d_interior}
\end{figure*}

The assumption of a pure carbon layer is a simplification, but a justifiable one for two reasons. The first reason is that our interior model solutions have enough flexibility (0--100\% carbon allowed for $100\times$ solar metallicity cases, see Figure \ref{fig:toi270d_interior}--\ref{fig:k218b_interior}) to account for EOS changes. In reality, carbon-rich planets forming outside of the soot line would accrete refractory organic materials that are rich in carbon but not purely carbon. \cite{bergin_exoplanet_2023} assumed a soot composition of C$_{100}$H$_{75-79}$O$_{11-17}$N$_{3-4}$S$_{1-3}$. The density of this organic soot is lower than that of pure carbon due to the inclusion of H, which will lead to lower $x_{\rm carbon}$ and higher $x_{\rm core}$ for the interior models. Given that $x_{\rm carbon}$ as high as 100\% is compatible with sub-Neptune M-R to within $1\sigma$ uncertainty, we do not expect that switching to soot EOS will change our results significantly: the physically plausible parameter space with 12--24 wt\% carbon-rich materials will likely remain as viable interior solutions. 

The second justification for a pure carbon layer is that high-pressure experiments favor the formation of a graphite (and diamond provided sufficient pressure) layer under carbon-saturated conditions. By studying carbon-enriched iron-silicate-carbon systems under high P-T conditions (1--2 GPa and 1523--1823 K), \cite{hakim_mineralogy_2019} found that the formation of a graphite layer on top of the silicate layer is favored in systems saturated with carbon. The saturation limit for carbon is low: only 0.05--0.7 wt\% of carbon is needed, significantly lower than the 12--24 wt\% range we consider here assuming efficient volatile retention. Graphite is also known to form and float to the surface in highly reduced magma oceans, creating a carbon-rich surface layer while limiting carbon sequestration into the core, for small planetary bodies with radii ranging from 25 km to $1\,R_\oplus$ \citep{h_keppler_graphite_2019}. Heat from formation and thermal blanketing by the H/He-rich atmosphere are expected to sustain long-lived magma oceans under sub-Neptune envelopes \citep[e.g.,][]{vazan_contribution_2018, kite_atmosphere_2020}. Hence, we expect soot-rich sub-Neptunes to have extremely reduced magma oceans which will lead to graphite flotation and the formation of a carbon layer on top of the silicate mantle. Note that, however, this assumption is based on extrapolation from smaller planetary bodies. To date, experimental evidence supporting graphite flotation on sub-Neptune-sized bodies is still lacking.

While by default, \texttt{CORGI} assumes an H/He envelope with solar metallicity, JWST observations of sub-Neptunes suggested that sub-Neptunes have metal-rich atmospheres \citep[e.g.,][]{gao_hazy_2023, benneke_jwst_2024, wogan_jwst_2024}. Therefore, in addition to the $1\times$ solar metallicity reference scenario, we further assume $100\times$ and $1000\times$ solar metallicities.


\subsection{Atmospheric Chemistry Model} \label{section:method_carbon_chem}
To simulate atmospheric chemistry of sub-Neptunes, we feed the atmospheric P-T profiles computed by \texttt{CORGI} into \texttt{Photochem} \citep{wogan_origin--life_2023}. Lower boundary conditions for the atmospheric chemistry model are calculated by a thermochemical equilibrium routine in \texttt{Photochem}. In the lower atmosphere where pressure and temperature are sufficiently high ($P\gtrsim100$ bar, $T\gtrsim1500$ K for our sub-Neptune models), thermochemical equilibrium dominates, and the abundances of molecules solely depend on the P-T condition and the number densities of H, He, O, C, N, and S atoms. We assume the same atmospheric metallicity grid ($1\times$, $100\times$, and $1000\times$ solar) as the interior model. In the $1\times$ solar metallicity model, O, N, S atoms have solar abundances relative to H and He, while C atom abundance is scaled relative to O atoms based on the input C/O ratio. Metallicities higher than $1\times$ solar are implemented by scaling up the abundances of O, C, N, S atoms by $100\times$ or $1000\times$ while keeping H and He abundances the same. As inputs to the chemistry model, we assume a fiducial atmospheric C/O grid from 0.4 to 2.0.

We choose this fiducial C/O grid by making the simplified assumption that the atmospheric layer has the same C/O ratio as the solid part of the planet (i.e., iron core, silicate mantle, and carbon layer). We assume the iron core is consisted of pure Fe, so do not contribute to the bulk solid C/O ratio. We assume that the silicate layer is consisted of pure MgSiO$_3$, i.e., Mg/Si=1 and Si/O=1/3. The Mg/Si ratios of stars in the solar neighborhood are indeed distributed around unity, according to the Hypatia catalog \citep{hinkel_stellar_2014}. Note that an Si/O ratio of 1/3 is greater than the solar Si/O ratio, which is approximately 1/15 \citep{asplund_chemical_2009}, but such a low oxygen content is consistent with the hypothetical origin of carbon-rich planets in an oxygen depleted region. Finally, the carbon layer is assumed to be consisted of pure C. Therefore, the overall C/O ratio of the solid part of the planet equals the ratio of C in the carbon layer to the O in the silicate layer. Following \cite{bergin_exoplanet_2023}, we assume an initial soot mass fraction of 12--24 wt\% before volatile loss. Using these two limits, the corresponding C/O ratios for these two compositions are roughly 0.57 and 1.32. As we will soon discuss, multiple mechanisms can drive the atmospheric C/O ratio to deviate from the planetary bulk ratio. Therefore, to account for possible carbon enrichment or depletion in the atmosphere, we assume the following fiducial C/O grid: 0.4, 0.5, 0.6, 0.8, 1.0, 1.2, 1.5, and 2.0.

By assuming that the atmospheric C/O ratio equals the interior C/O ratio, we are essentially assuming that atmospheric evolution does not cause the atmospheric C/O of sub-Neptunes to deviate from the primordial building blocks. Extreme ultraviolet radiation and X-rays, collectively known as XUV, can drive efficient atmospheric escape on exoplanets, especially when the host stars are young and XUV-active \citep[e.g.,][]{tian_thermal_2009, owen_kepler_2013, lopez_understanding_2014, luger_extreme_2015, zahnle_cosmic_2017}. M dwarfs, the host stars of our three prototypical sub-Neptunes, are particularly active in XUV. A recent study found that cumulative atmospheric loss due to XUV may yield more small planets around M dwarfs airless than predicted from canonical theories \citep{pass_receding_2025}. The effects of XUV-driven escape on the atmospheric compositions of hypothetical carbon-rich sub-Neptunes, therefore, must be taken into consideration. 

The most important effect of XUV-driven escape is the increase in metallicity. Under intense XUV irradiation, atmospheric escape on exoplanets orbiting M dwarfs mainly occur via energy-limited hydrodynamic wind \citep{luger_extreme_2015, sengupta_upper_2016}. Hydrogen dominates the energy-limited escape flux due to its light weight, leading to an increase of the heavy atom to hydrogen ratio, i.e., metallicity. Our model naturally handles this effect by including two high metallicity ($100\times$ and $1000\times$ solar) cases.

Heavy atoms like O and C can be entrained in the hydrodynamic outflow and escape to space as well. The upward flux of heavy atoms depend on mixing ratio and atomic mass, so in principle, mass fractionation can occur as a result of energy-limited escape, where lighter atoms like C will escape more efficiently than heavier atoms like O \citep{hunten_mass_1987}, leading to a decreased C/O ratio despite a carbon-rich interior. However, to date no observational constraints on the extent of C/O mass fractionation during escape exists for sub-Neptunes. Even though escaping O and C features have been detected on hot Jupiters \citep[e.g.,][]{koskinen_escape_2013, koskinen_escape_2013-1}, the exact C/O ratio of the outflow is not constrained. Therefore, while we acknowledge the caveat that our assumption that atmospheric C/O equals interior C/O may be affected by atmospheric escape, precise constraints on post-escape C/O ratios are not available. The lowest C/O ratio in our C/O grid, a subsolar value of 0.4, can be considered as a pessimistic lower limit for carbon retention.

In the upper atmosphere accessible to remote observation, photochemistry dictates the composition of the atmosphere, and this processed is also modeled using \texttt{Photochem} \citep{wogan_origin--life_2023}. Photochemistry in sub-Neptune atmospheres are driven by high-energy photons from the host stars, so the choice of input stellar spectra is critical. The stellar spectrum of GJ 1214 was observed by the MUSCLES spectral survey \citep{france_muscles_2016, youngblood_muscles_2016, loyd_muscles_2016}, which we employ as the input to \texttt{Photochem}. K2-18 and TOI-270, however, lack such observations covering ultraviolet wavelengths. Therefore, we employ the MUSCLES spectrum for GJ 436 as an approximation, due to its similarity to K2-18 and TOI-270 in terms of effective temperature and size. All input stellar spectra are scaled to the flux received by the planets. 
Note that the C/O ratio in the upper atmosphere accessible to spectroscopic observations does not necessarily equal the bulk atmosphere or bulk planet C/O ratio. All C/O ratios we quote in results below are bulk atmosphere C/O ratios, which are assumed to equal the interior C/O ratios.

The deep atmosphere dominated by thermochemical equilibrium and the shallow atmosphere dominated by photochemistry are connected by vertical transport. Vertical transport in the atmosphere is separated into two regimes. For the lower atmosphere ($P>5$ bar), we assume a constant eddy diffusion coefficient of $K_{zz, 0} = 6\times10^8$ cm$^2$ s$^{-1}$ following \cite{bergin_exoplanet_2023}. For the upper atmosphere ($P<5$ bar), we assume an inverse square-root dependence on pressure \citep{tsai_photochemically_2023}, $K_{\rm zz}(P) = K_{\rm{zz}, 0} (5/P)^{0.5}$, where $P$ is in the unit of bar and $K_{\rm zz}$ in the unit of cm$^2$ s$^{-1}$.

\subsection{Transmission Spectra Model} \label{section:method_carbon_spectra}
To simulate the transmission spectra of carbon-rich sub-Neptunes, we use the open-source exoplanet spectra simulation code \texttt{petitRADTRANS} \citep{molliere_petitradtrans_2019, blain_spectralmodel_2024}. Hydrocarbon hazes are expected to form in the atmospheres of carbon-rich sub-Neptunes \citep{bergin_exoplanet_2023}. Clouds form more easily in metal-rich atmospheres \citep{hoyer_extremely_2023}. Therefore, we assume the following haze and cloud conditions: clear, $10\times$ haze factor (i.e., scaling Rayleigh scattering opacities by $10\times$) without clouds, and $10\times$ haze factor with a gray cloud deck at 10 mbar. 

We perform $\chi^2$ tests on all simulated spectra to calculate if they are consistent with JWST observations. HST observations are also shown for visual comparison, but are not incorporated into the $\chi^2$ tests, because JWST data with higher signal-to-noise ratio covering a wider wavelength range are available.
For TOI-270 d, we include HST/WFC3 \citep{mikal-evans_hubble_2023}, JWST/NIRISS \citep{benneke_jwst_2024}, and JWST/NIRSpec \citep{holmberg_possible_2024} observations.
For GJ 1214 b, we include HST/WFC3 \citep{kreidberg_clouds_2014}, JWST/NIRSpec \citep{schlawin_possible_2024}, and JWST/MIRI \citep{kempton_reflective_2023} observations.
For K2-18 b, we include the HST/WFC3 observation by \cite{benneke_water_2019} and the JWST/NIRISS and JWST/NIRSpec observations by \cite{madhusudhan_carbon-bearing_2023}.

Offsets are needed to account for baseline shifts between different instruments and the two NIRSpec detectors, NRS1 and NRS2 \citep[see e.g.,][]{madhusudhan_carbon-bearing_2023, schlawin_possible_2024, schmidt_comprehensive_2025}. When performing $\chi^2$ tests, we divide simulated spectra into segments based on instrument/detector wavelength ranges and allow each segment to shift vertically.

\section{Results} \label{section:results_carbon}
Here, we report our results for the interior compositions (Section \ref{sec:result_carbon_interior}) and transmission spectra (Section \ref{sec:result_carbon_spectra}) of carbon-rich sub-Neptunes.

\subsection{Carbon-rich Interiors Are Consistent with Mass-Radius Measurements} \label{sec:result_carbon_interior}
The measured masses and radii of the three prototypical sub-Neptunes we consider are all consistent with carbon-rich compositions. For GJ 1214 b, however, a carbon-rich interior is inconsistent with its atmospheric metallicity. Here, we discuss each planet separately.

\subsubsection{TOI-270 d}
Our results show that TOI-270 d can have a carbon-rich interior composition with any amount of carbon if the atmospheric metallicity is $1\times$ or $100\times$ solar, but only physically implausible extremely carbon-rich interior compositions exist for the $1000\times$ solar metallicity scenario (Figure \ref{fig:toi270d_interior}), under the assumption of a four-layer interior structure with an iron core, a silicate mantle, a carbon layer, and an atmosphere.

Sub-Neptune TOI-270 d has a mass of $4.78\pm0.43\,M_\oplus$ and a radius of $2.133\pm0.058\,R_\oplus$
\citep{van_eylen_masses_2021}.
If TOI-270 d has a $1\times$ solar metallicity atmosphere, any mass fraction of carbon is allowed, while the envelope mass fraction $x_{\rm env}$ is limited to $\lesssim0.4\%$ (Figure \ref{fig:toi270d_interior}a). 
If TOI-270 d has a $100\times$ solar metallicity atmosphere, it is still allowed to have any mass fraction of carbon, while the atmosphere mass fraction is limited to $\lesssim10\%$ (Figure \ref{fig:toi270d_interior}b). Note that \cite{benneke_jwst_2024} retrieved an atmospheric metallicity that is broadly consistent with $100\times$ solar ($224.6^{+98.1}_{-86.1}\times$ solar).
If the atmospheric metallicity of TOI-270 d is as high as $1000\times$ solar, no compositional solution is consistent within the $1\sigma$ M-R uncertainties. Only implausible cases with $\gtrsim40\%$ $x_{\rm carbon}$ can match the observed M-R within $2\sigma$ or $3\sigma$ uncertainties (Figure \ref{fig:toi270d_interior}c).

\subsubsection{GJ 1214 b}
\begin{figure}[t!]
\centering
\includegraphics[width=0.85\columnwidth]{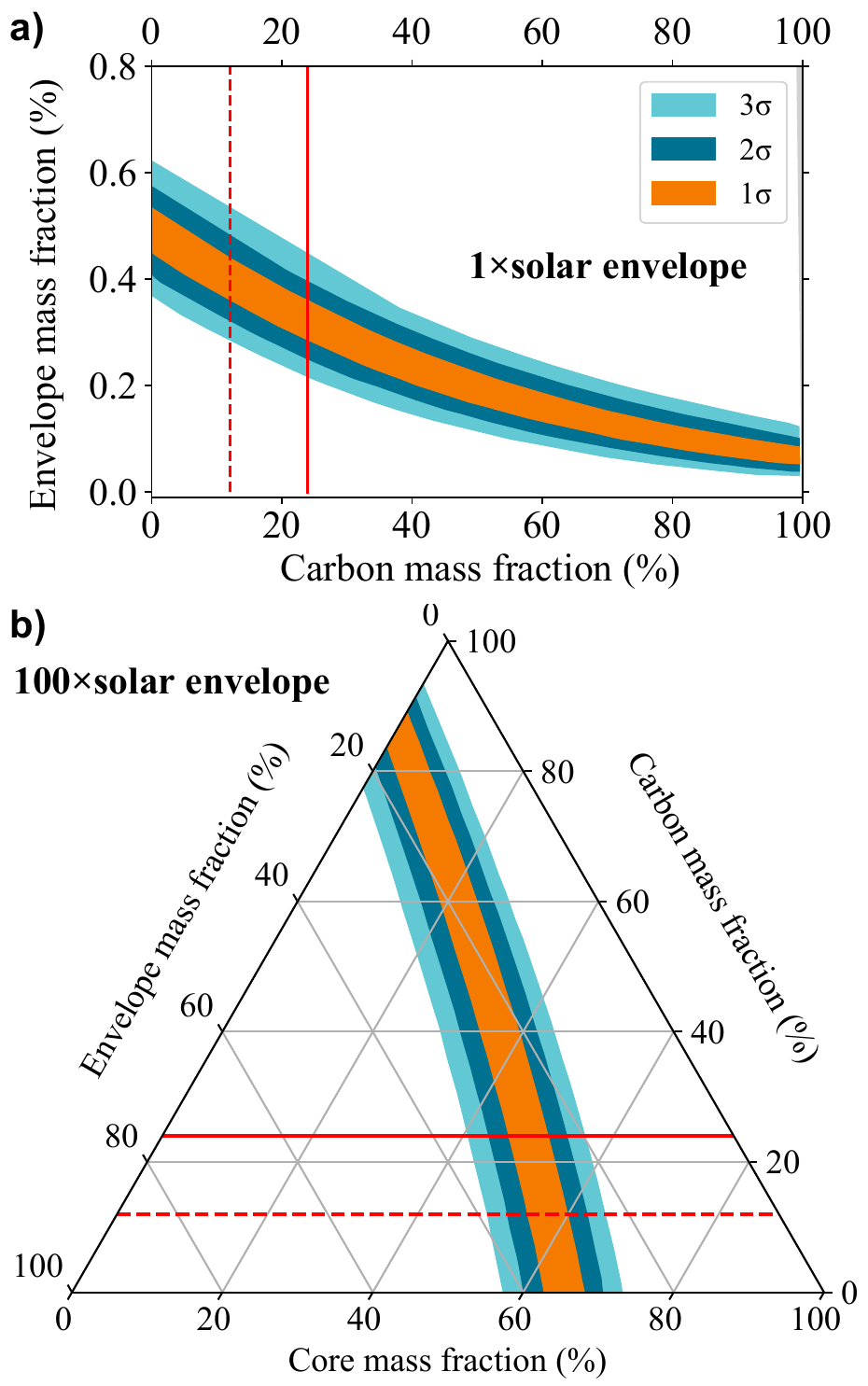}
\caption{Possible interior compositions of GJ 1214 b. Compositions consistent within the $1\sigma$, $2\sigma$, and $3\sigma$ M-R uncertainties are shown in orange, dark teal, and light teal, respectively. The red dashed and solid lines show $x_{\rm carbon}=12$ wt\% and 24 wt\%, respectively, the upper bounds of $x_{\rm carbon}$ assuming no volatile loss \citep{bergin_exoplanet_2023}. $x_{\rm carbon}$ defines the bulk carbon content of the planet and in turn defines the C/O ratio in the atmosphere, under the assumption that the atmospheric C/O ratio is the same as the bulk solid C/O ratio. (a) Carbon-envelope diagram for $1\times$ solar metallicity envelope. (b) Ternary diagram for $100\times$ solar metallicity envelope. The lack of $1000\times$ solar metallicity solution rules out carbon-rich interiors for GJ 1214 b, because previous works favored high metallicities (see text).}
\label{fig:gj1214b_interior}
\end{figure}

Our interior model results rule out a carbon-rich interior for GJ 1214 b, under the assumption of a four-layer interior structure with an iron core, a silicate mantle, a carbon layer, and an atmosphere. Solely considering interior models, GJ 1214 b can have a carbon-rich interior with arbitrary $x_{\rm carbon}$, assuming $1\times$ or $100\times$ solar metallicity (Figure \ref{fig:gj1214b_interior}). No solution exist at all for a three-layer core-carbon-envelope structure if the atmospheric metallicity is $1000\times$ solar.

An extremely metal-rich atmosphere, however, is strongly favored for GJ 1214 b. Previous photochemical haze models largely ruled out $\leq300\times$ solar metallicity and favored an extremely high metallicity of $\geq1000\times$ solar \citep{gao_hazy_2023}. \cite{kempton_reflective_2023} even considered a metallicity of $3000\times$ solar. If GJ 1214 b atmosphere indeed has a $\geq1000\times$ metallicity, it is impossible that GJ 1214 b has a carbon-rich composition according to our interior model, unless the Fe/Si ratio of the core is substantially altered.

Note that a carbon-rich interior for GJ 1214 b is ruled out because of the high density of $1000\times$ solar metallicity envelope, but do not rely on our assumption that the atmospheric C/O ratio equals the interior C/O ratio. Tuning the atmospheric C/O ratio produces at max 25\% difference (i.e., the difference between C and O atomic mass) in the mean molecular weight, while increasing atmospheric metallicity from $100\times$ to $1000\times$ increases the mean molecular weight by threefold. Hence, metallicity has a much greater effect than C/O on the planet's radius given an atmosphere with a certain mass fraction, which plays a major role in the interior model.

A caveat of our result for GJ 1214 b is that due to the extremely high metallicity, the atmosphere of GJ 1214 b may be secondary \citep{gao_hazy_2023}, despite a controversial detection of escaping helium from GJ 1214 b may rule out this possibility \citep{kasper_nondetection_2020, spake_non-detection_2022, orell-miquel_tentative_2022}. If the atmosphere of GJ 1214 b is indeed secondary, our assumption that atmospheric C/O ratio equals that of the interior, which naturally implies a primordial origin, is challenged and our claim that GJ 1214 b is not a carbon-rich sub-Neptune needs to be reassessed.

\subsubsection{K2-18 b}
\begin{figure*}[ht!]
\centering
\includegraphics[width=\textwidth]{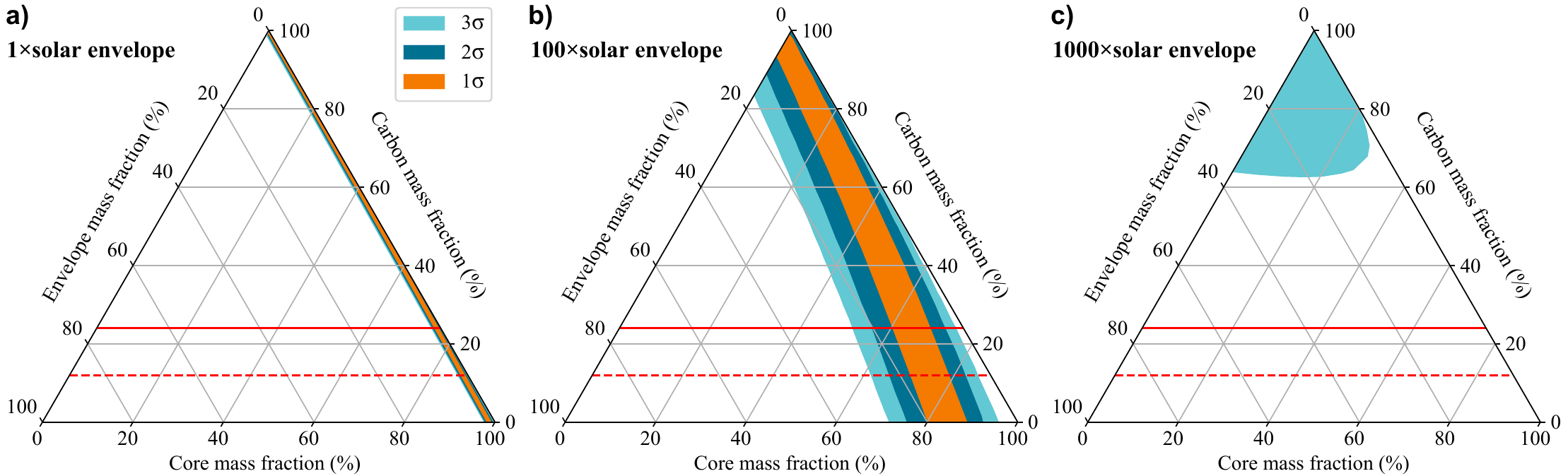}
\caption{Possible interior compositions of K2-18 b. Compositions consistent within the $1\sigma$, $2\sigma$, and $3\sigma$ M-R uncertainties are shown in orange, dark teal, and light teal, respectively. The red dashed and solid lines show $x_{\rm carbon}=12$ wt\% and 24 wt\%, respectively, the upper bounds of $x_{\rm carbon}$ assuming no volatile loss \citep{bergin_exoplanet_2023}. $x_{\rm carbon}$ defines the bulk carbon content of the planet and in turn defines the C/O ratio in the atmosphere, under the assumption that the atmospheric C/O ratio is the same as the bulk solid C/O ratio. Ternary diagrams are shown for (a) $1\times$ solar metallicity, (b) $100\times$ solar metallicity, and (c) $1000\times$ solar metallicity scenarios. The $100\times$ solar metallicity scenario allows the widest range of interior compositions and is consistent with previous modeling and retrieval works \citep{wogan_jwst_2024, schmidt_comprehensive_2025}.}
\label{fig:k218b_interior}
\end{figure*}

Our results show that K2-18 b is consistent with carbon-rich interior compositions if the atmospheric metallicity is $1\times$ or $100\times$ solar, but the $1000\times$ solar metallicity scenario only produces physically implausible solutions (Figure \ref{fig:k218b_interior}), similar to TOI-270 d, under the assumption of a four-layer interior structure with an iron core, a silicate mantle, a carbon layer, and an atmosphere.

Sub-Neptune K2-18 b has a mass of $8.63\pm1.35\,M_\oplus$ and a radius of $2.610\pm0.087\,R_\oplus$
\citep{benneke_water_2019}. Arbitrary combinations of $x_{\rm core}$ and $x_{\rm carbon}$ are allowed for the $1\times$ solar case, while $x_{\rm env}$ is limited to $\lesssim3\%$ (Figure \ref{fig:k218b_interior}a). A $100\times$ solar assumption relaxes the constraints on $x_{\rm env}$ to $\lesssim30\%$ (Figure \ref{fig:k218b_interior}b). Under the assumption of $1000\times$ solar envelope, only the implausible high-carbon corner is consistent with the $3\sigma$ M-R uncertainties of K2-18 b (Figure \ref{fig:k218b_interior}c). Previous modeling works suggested that a gas-rich sub-Neptune composition with a $100\times$ solar metallicity envelope is compatible with JWST observations \citep{wogan_jwst_2024, schmidt_comprehensive_2025}.
We discuss the consistency with HST and JWST observations of our $100\times$ solar metallicity carbon-rich interior models in the next section.

\subsection{Carbon-rich Interiors Are Consistent with HST and JWST Observations} \label{sec:result_carbon_spectra}
In this section, we show that carbon-rich interiors produce spectra that are compatible with HST and JWST observations.
The chemical profiles are shown in Appendix \ref{sec:app_chemical_profile}.
We do not discuss $1\times$ solar metallicity spectra here because their large atmospheric scale heights are totally inconsistent with HST and JWST observations that suggest metal-rich atmospheres.

\begin{figure*}[ht!]
\centering
\includegraphics[width=0.95\textwidth]{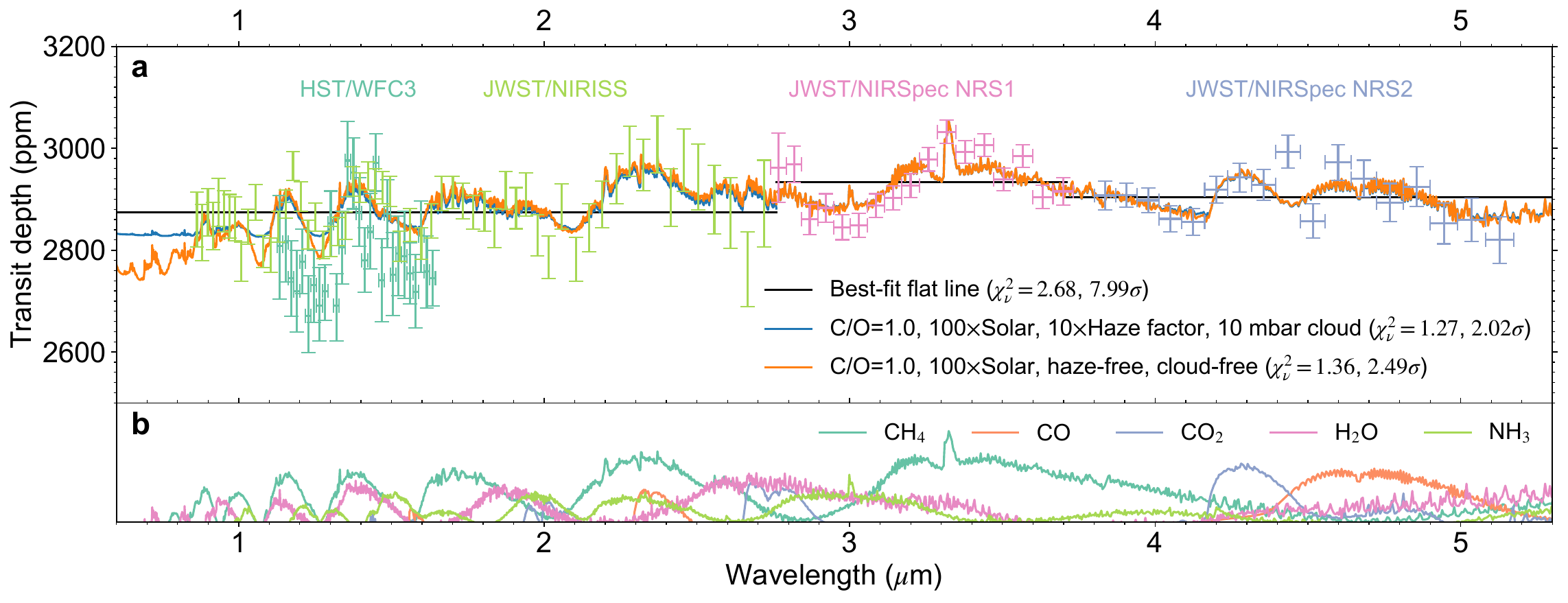}
\caption{(a) Simulated transmission spectra of sub-Neptune TOI-270 d compared with HST and JWST observations. The best-fit flat line, best-fit hazy model, and best-fit haze-free model are plotted, with their respective reduced $\chi_\nu^2$ and $\sigma$ values shown. Note that the HST/WFC3 data are shown for comparison, but the models are only fitted to the JWST data. The simulated spectra are allowed to shift to account for baseline offsets between the different instruments/detectors. 
(b) Contributions to the overall transmission spectra by the individual molecules. A carbon-rich interior composition with a $100\times$ solar metallicity envelope is compatible with HST and JWST transmission spectra observations of TOI-270 d.}
\label{fig:toi270d_spectra}
\end{figure*}

\subsubsection{TOI-270 d}
A carbon-rich interior composition with a $100\times$ solar metallicity envelope is consistent with the JWST transmission spectra of TOI-270 d (within $3\sigma$ of observations according to our $\chi^2$ tests, Figure \ref{fig:toi270d_spectra}). The best-fit carbon-rich atmosphere model has a C/O ratio of 1.0, corresponding to a 19.3 wt\% carbon layer relative to the solid part of the planet, assuming that the atmospheric C/O ratio is the same as the C/O ratio of the solid part of the planet. A $\sim5$ wt\% atmosphere is allowed by our interior model given this relative carbon layer mass fraction (Figure \ref{fig:toi270d_interior}b), so the $x_{\rm carbon}$ relative to the bulk planetary mass in this case is roughly 18.3 wt\%. This result is robust over a range of haze parameters (from haze-free to $10\times$ haze factor) and cloud conditions (from cloud-free to 10 mbar gray cloud deck). Several other models with C/O ratios equal 0.8 and 1.2 also passed the $\chi^2$ test, indicating that a wide range of carbon mass fractions are allowed for the interior of TOI-270 d. These C/O ratios translate into 16.1 wt\% to 22.4 wt\% carbon layers relative to the solid part of the planet, respectively, assuming atmospheric C/O equals bulk interior C/O. When considering a $\sim5$ wt\% atmosphere (Figure \ref{fig:toi270d_interior}b), these numbers shrink to 15.29 wt\% and 21.28 wt\% carbon relative to the total planet mass. The above results are significant because it offers another observationally consistent interior structure for sub-Neptunes. In addition to water world compositions \citep{rigby_ocean_2024, holmberg_possible_2024}, Neptune-like ice-rich compositions \citep{leconte_3d_2024, wogan_jwst_2024}, and ice-poor compositions with substantial amount of iron and silicates \citep{benneke_jwst_2024} that have been suggested previously, the presence of a carbon layer indicative of formation between the refractory organic carbon soot line and the water ice line \citep{bergin_exoplanet_2023} is another plausible alternative.

In Figure \ref{fig:toi270d_spectra}, we show the best-fit flat line, hazy, and haze-free models for TOI-270 d. The latter two models are compatible with JWST observations (within $3\sigma$), while the flat line model is rejected at $7.99\sigma$. Dominant features in our model spectra agree well with \cite{holmberg_possible_2024} and \cite{benneke_jwst_2024}: CH$_4$ features at 1.15, 1.4, 1.7, 2.3, and 3.3 $\mu$m, H$_2$O features at 1.4 and 1.9 $\mu$m, and CO$_2$ features at 4.2 $\mu$m. A noticeable difference is that in our models, CO shows the fundamental 1-0 band from 4.5--5.0 $\mu$m, matching a bump in the NIRSpec wavelengths, while \cite{holmberg_possible_2024} and \cite{benneke_jwst_2024} attributed this bump to CS$_2$. This difference is due to the more reduced chemical environment of a carbon-rich interior. CO concentration in our model, $\log(\textrm{CO})=-1.26$, is higher than the upper bounds placed by previous works. Hence, we do not require sulfur species (CS$_2$ and SO$_2$) to improve the fit.


\subsubsection{GJ 1214 b}
Our simulated transmission spectra (Appendix \ref{sec:gj1214b_spectra}) further support that a carbon-rich interior composition can be ruled out for GJ 1214 b, under the assumption of a four-layer interior structure with an iron core, a silicate mantle, a carbon layer, and an atmosphere. Because of the flatness of the transmission spectra of GJ 1214 b, all of our simulated spectra with $1\times$ or $100\times$ solar metallicity are firmly rejected by the $\chi^2$ test ($p$-value $\ll1$), regardless of the C/O ratio. Extremely metal-rich ($\geq1000\times$ solar) and hazy atmospheres may produce compatible spectra, but such high metallicities are ruled out by our interior model.
Therefore, no viable carbon-rich solution exists for GJ 1214 b.
Note that, however, this argument is based purely on the metallicity of the atmosphere but does not rely on our assumption that atmospheric C/O equals interior C/O, because scaling atmospheric C/O has much smaller effect on the mean molecular weight than increasing the atmospheric metallicity from $100\times$ to $1000\times$ solar.

\begin{figure*}[ht!]
\centering
\includegraphics[width=0.95\textwidth]{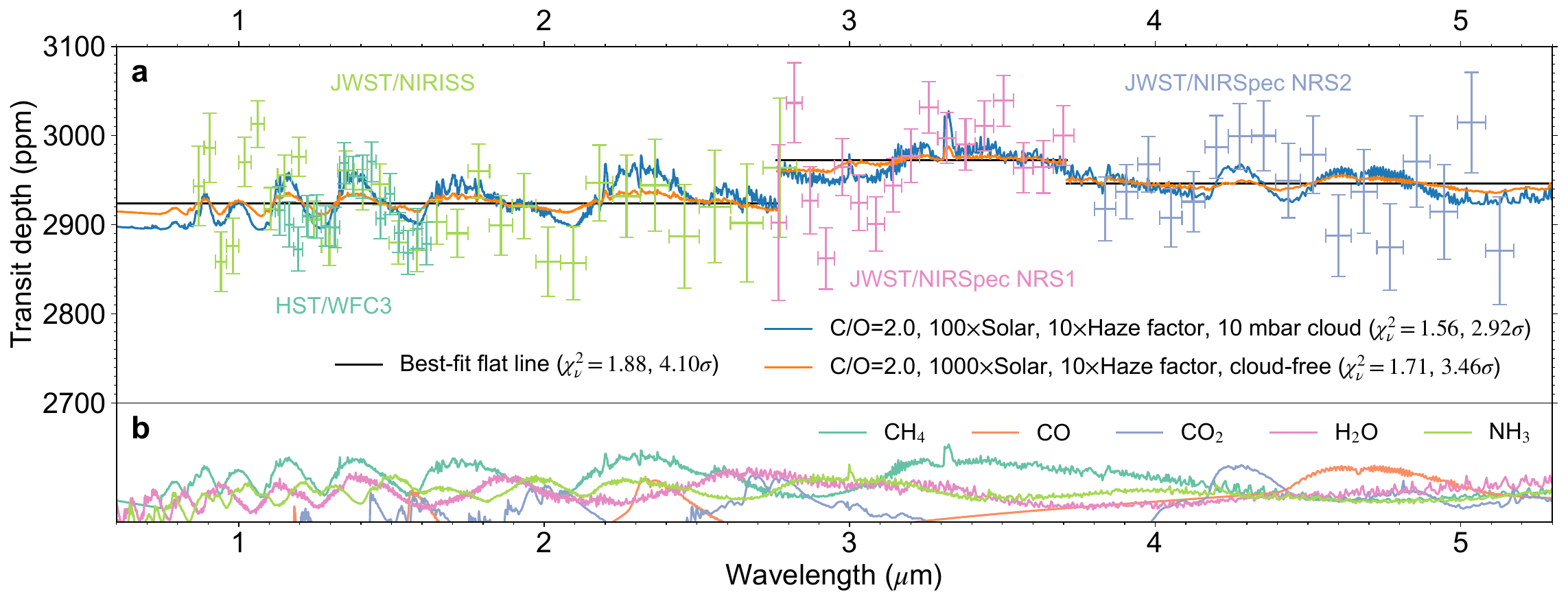}
\caption{(a) Simulated transmission spectra of sub-Neptune K2-18 b compared with HST and JWST observations. The best-fit flat line model, best-fit $100\times$ solar metallicity model, and the best-fit $1000\times$ solar metallicity model are shown. Note that the HST/WFC3 data are shown for comparison, but the models are only fitted to the JWST data. 
(b) Contributions to the overall transmission spectra by the individual molecules. The best-fit $100\times$ solar metallicity model is barely consistent with the data, while all other models are ruled out at $>3\sigma$. A high C/O ratio of 2.0 is favored, consistent with the retrieval work by \cite{schmidt_comprehensive_2025}, which constrained the mixing ratio of CH$_4$ to be $\log(\textrm{CH}_4)=-1.15^{+0.40}_{-0.52}$. Additional CH$_4$ and CO$_2$ sources are needed to fully match the JWST observations.}
\label{fig:k218b_spectra}
\end{figure*}

\subsubsection{K2-18 b} \label{sec:result_k218b_atm}
A carbon-rich interior composition is broadly compatible with the HST and JWST transmission spectra of K2-18 b (Figure \ref{fig:k218b_spectra}). The best-fit $100\times$ solar metallicity model is barely within $3\sigma$ of the JWST observations at $2.92\sigma$, while the best-fit flat line is ruled out at $4.1\sigma$. The best-fit carbon-rich atmosphere model for K2-18 b has C/O=2.0. Under the assumption that atmospheric C/O equals interior C/O, this corresponds to a carbon fraction of 32.4 wt\% relative to the solid part of the planet. Given that according to our interior model, $\sim10$\% of K2-18's mass is in the atmosphere, assuming a $100\times$ solar metallicity, $x_{\rm carbon}$ relative to the bulk planet mass is scaled to 29.2 wt\%. This high carbon layer mass fraction, however, is above the 24 wt\% upper limit assuming no volatile loss \citep{bergin_exoplanet_2023}.

Therefore, even though the transmission spectra of K2-18 b is broadly compatible with a carbon-rich interior, additional atmospheric carbon sources are required to strengthen this compatibility, while keeping the interior $x_{\rm carbon}$ reasonably low. The key feature of the spectrum of K2-18 b is the enrichment of carbon species: the log concentration of CH$_4$ is estimated to be $-1.15^{+0.40}_{-0.52}$ \citep{schmidt_comprehensive_2025} and the log concentration of CO$_2$ is as high as $-2.09^{+0.51}_{-0.94}$ \citep{madhusudhan_carbon-bearing_2023}. The CH$_4$ and CO$_2$ mixing ratios in our models are slightly lower: $\log(\textrm{CH}_4)=-1.28$ (which is within error of \citealt{schmidt_comprehensive_2025}) and $\log(\textrm{CO}_2)=-3.39$. Increasing the concentration of carbon species in our models will lead to better fit for CH$_4$ features at 1.15, 1.4, 1.7, 2.3, and 3.3 $\mu$m, and the CO$_2$ feature at 4.2 $\mu$m. We discuss the potential sources of additional carbon species in the next section.

\section{Discussion and Conclusion} \label{section:discussion_carbon}
In this paper, we use a combination of planetary interior, atmospheric chemistry, and transmission spectra models to investigate whether carbon-rich compositions are compatible with HST and JWST observations of sub-Neptunes. Our results for prototypical sub-Neptunes TOI-270 d and K2-18 b have validated that carbon-rich interiors, which are capable of producing high C/O ($\geq1.0$) and metal-rich ($\sim100\times$ solar metallicity) atmospheres, can produce spectral signatures consistent with HST and JWST observations, under the assumption that atmospheric C/O ratio equals the interior C/O ratio. Therefore, a new category of interior composition consisting of a iron-silicate core, a carbon layer, and an H/He-rich envelope can be added to the existing list of possible interior compositions of sub-Neptunes. In addition, our results for GJ 1214 b have shown that a combination of interior and atmosphere models can firmly rule out carbon-rich interior composition, thereby reducing the interior degeneracy of this planet, under the assumption of a four-layer interior structure with an iron core, a silicate mantle, a carbon layer, and a primordial atmosphere that reflects the C/O ratio of the interior.

This exploratory work makes a simplification by ignoring chemical interactions between the carbon layer and the atmosphere. Sub-Neptunes may have long-lived magma oceans that actively react with the atmosphere \citep[e.g.,][]{vazan_contribution_2018, kite_atmosphere_2020, schlichting_chemical_2022, misener_atmospheres_2023}. Redox state of the magma ocean have impacts on atmospheric composition \citep{gaillard_redox_2022}. Even without the presence of a magma ocean, carbon-hydrogen chemical reactions are expected to occur. \cite{pena-alvarez_-situ_2021} experimentally investigated into the reaction between H$_2$ and diamond, graphite, and disordered carbonaceous materials. They found that under high P-T conditions (0.5--3 GPa and $\sim600$--800 K), CH$_4$ and other hydrocarbons such as C$_2$H$_6$ are formed. The high P-T reactions between carbon and hydrogen provides a potential lower boundary source for reducing carbon species, possibly explaining the CH$_4$ enrichment in K2-18 b (Section \ref{sec:result_k218b_atm}).

Several other high-pressure experimental studies may also have implications on the carbon-rich sub-Neptunes hypothesized in this work. \cite{benedetti_dissociation_1999} showed that CH$_4$ turns into diamond at 10--50 GPa and $\sim$2000--3000 K, a P-T condition relevant for the deep atmosphere of carbon-rich sub-Neptunes. Similarly, formation of diamond from CH$_4$ was experimentally observed by \cite{hirai_polymerization_2009} at 10--80 GPa and $>3000$ K, a broadly consistent P-T condition. The dissociation of methane into diamond may serve as a sink for atmospheric carbon if diamond rainout into the interior occurs, reducing the atmospheric C/O ratio. Diamond production from hydrocarbons heavier than CH$_4$ was also observed when \cite{kraus_formation_2017} experimentally compressed (C$_8$H$_8$)$_n$ to roughly 150 GPa and 5000 K. However, a recent study by \cite{tabak_evidence_2024} found that CH$_4$ does not fully dissociate into diamond between 10--50 GPa, but remain stable until transitioning into polymers and subsequently dissociating at 80--150 GPa. \cite{kadobayashi_diamond_2021} investigated the effect of adding oxygen to hydrocarbons under high pressure and temperature, and found that diamonds can form in the C-O-H system under milder P-T conditions (45 GPa and $\sim1600$ K) than in a purely C-H environment. Therefore, the correlation between carbon layer mass and the atmospheric redox state is complicated and potentially counterintuitive: a planet with an O-rich atmosphere may still have a diamond-rich interior due to hydrocarbon dissociation. Future experimental works on the interior-atmosphere interactions in carbon-rich sub-Neptunes will offer better insights into such planets than our simplified assumption that atmospheric C/O ratio is the same as the C/O ratio of the solid part of the planet.

In summary, carbon-rich composition is an observationally compatible alternative explanation to the interiors of sub-Neptunes. We call for high-pressure experiments for the EOS of refractory organic carbon and carbon-hydrogen chemical reactions. We also call for formation and evolution models investigating whether sub-Neptunes can be formed between the soot line and the water snow line, which is required for carbon-rich interiors without host star C/O enrichment.


\begin{acknowledgments}
We thank the anonymous reviewer for insightful comments that improved the quality of the manuscript. Z.L. acknowledges funding from the Center for Matter at Atomic Pressures (CMAP), a National Science Foundation (NSF) Physics Frontiers Center, under award PHY-2020249. The authors acknowledge the MIT SuperCloud and Lincoln Laboratory Supercomputing Center for providing high performance computing resources that have contributed to the research results reported within this paper.
\end{acknowledgments}

\begin{contribution}
Z.L. designed the study, generated the models, and wrote the manuscript. S.S. supervised and edited the manuscript. 


\end{contribution}

%

\software{\texttt{CORGI} \citep{lin_interior_2025}, \texttt{petitRADTRANS} \citep{molliere_petitradtrans_2019, blain_spectralmodel_2024}, \texttt{Photochem} \citep{wogan_origin--life_2023}, \texttt{Matplotlib} \citep{Hunter_2007_matplotlib}, \texttt{mpltern} \citep{yuji_ikeda_yuzie007mpltern_2024}, \texttt{NumPy} \citep{harris2020array}, \texttt{SciPy} \citep{2020SciPy-NMeth}.}


\appendix


\section{Interior Model Intercomparison} \label{sec:app_model_validation}
Although the \texttt{CORGI} planetary interior model is thoroughly validated in \cite{lin_interior_2025}, EOSs for diamond and graphite are newly added for this work and have not been benchmarked. For model robustness, here we compare our interior models to the previous interior modeling work for 55 Cnc e by \cite{madhusudhan_possible_2012}.
 
In addition to carbon EOS, we also incorporate up-to-date SiC EOS in \texttt{CORGI}. SiC is the dominant form of Si condensing from carbon-rich (C/O $>1$) disks instead of silicate \citep[e.g.,][]{kuchner_extrasolar_2005, seager_massradius_2007, madhusudhan_possible_2012}. Even though in this work we do not include a SiC layer in our sub-Neptune interior models, SiC may still play a role at the silicate-carbon layer boundary. Therefore, for completeness, we add SiC EOS to \texttt{CORGI}.

We adopt the SiC EOS from \cite{miozzi_equation_2018} at low pressures ($<426$ GPa), which was fitted to experimental data using the Vinet EOS \citep{vinet_compressibility_1987, vinet_universal_1989} up to 200 GPa and extrapolated beyond. At high pressures ($>426$ GPa), we switch to a Vinet fit to the \cite{wilson_interior_2014} ab initio simulation because the extrapolation from experimental data begins to deviate from the simulations. 426 GPa is where the two Vinet fits intersect and hence chosen as a switching point. We assume that SiC transforms from a low-pressure zinc-blende structure to a high-pressure rock-salt structure at 67.5 GPa, following \cite{miozzi_equation_2018}.

\begin{figure*}[ht!]
\centering
\includegraphics[width=0.7\textwidth]{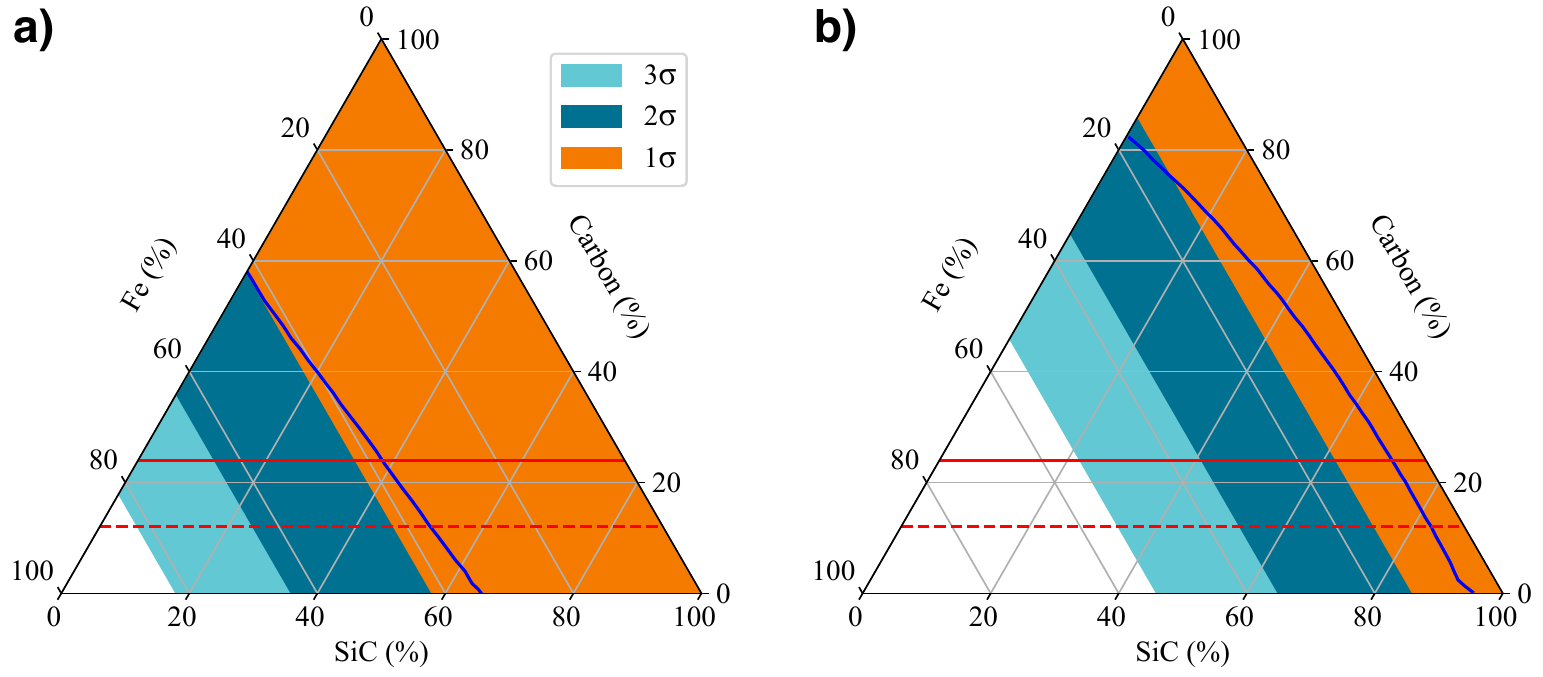}
\caption{Possible interior compositions of 55 Cnc e. Compositions consistent within the $1\sigma$, $2\sigma$, and $3\sigma$ M-R uncertainties are shown in orange, dark teal, and light teal, respectively. The red dashed and solid lines show $x_{\rm carbon}=12$ wt\% and 24 wt\%, respectively, the upper bounds of $x_{\rm carbon}$ assuming no volatile loss \citep{bergin_exoplanet_2023}. The blue curve shows interior solution from \cite{madhusudhan_possible_2012}, to the right of which are allowed values ($1\sigma$ uncertainty). Ternary diagrams are shown for two radius assumptions: (a) visible radius, $2.04\pm0.15\,R_\oplus$, and (b) gray radius, $2.20\pm0.12\,R_\oplus$, following \cite{madhusudhan_possible_2012}. Our interior models with updated carbon and SiC EOSs are consistent with earlier studies.}
\label{fig:55cnce_interior}
\end{figure*}

To check for consistency with previous models given updated carbon and SiC EOSs, we reproduce the interior models for 55 Cnc e in \cite{madhusudhan_possible_2012}. For carbon, \cite{madhusudhan_possible_2012} used the Birch-Murnagham graphite EOS from \cite{wang_crystal_2012}, the graphite-diamond phase transition from \cite{bundy_pressure-temperature_1996}, and the \cite{dewaele_high_2008} diamond EOS. For SiC, \cite{madhusudhan_possible_2012} adopted the Birch–Murnaghan EOS from \cite{seager_massradius_2007}. The planet 55 Cnc e is assumed to be consisted of three layers: an iron core, a SiC mantle, and a carbon layer. We assume a mass of $8.37\pm0.38\,M_\oplus$ \citep{endl_revisiting_2012} and two radii: visible radius $2.04\pm0.15\,R_\oplus$ \citep{winn_super-earth_2011, gillon_improved_2012} and gray radius $2.20\pm0.12\,R_\oplus$ \citep{demory_detection_2011, gillon_improved_2012}. Note that the M-R measurement of 55 Cnc e has been updated since the publication of \cite{madhusudhan_possible_2012}. The up-to-date values are $M_p = 7.99^{+0.32}_{-0.33}\,M_\oplus$ and $R_p = 1.875\pm0.029\,R_\oplus$ \citep{bourrier_55_2018}. However, we adopt the outdated M-R values above to ensure consistent model intercomparison. 55 Cnc e interior compositions calculated by \texttt{CORGI} are shown in Figure \ref{fig:55cnce_interior}. Agreements are observed between our models and \cite{madhusudhan_possible_2012} models. However, we note that this agreement does not imply that our models reflect the physical reality at high pressures, which can only be achieved by cross-validation against experimental measurements. Rather, we are showing that our theoretical model, which are subjected to the uncertainties of theoretical EOS or extrapolation from experimental data, behaves similarly as previous theoretical models.

\section{Chemical Profiles} \label{sec:app_chemical_profile}
\begin{figure*}[ht!]
\centering
\includegraphics[width=0.8\textwidth]{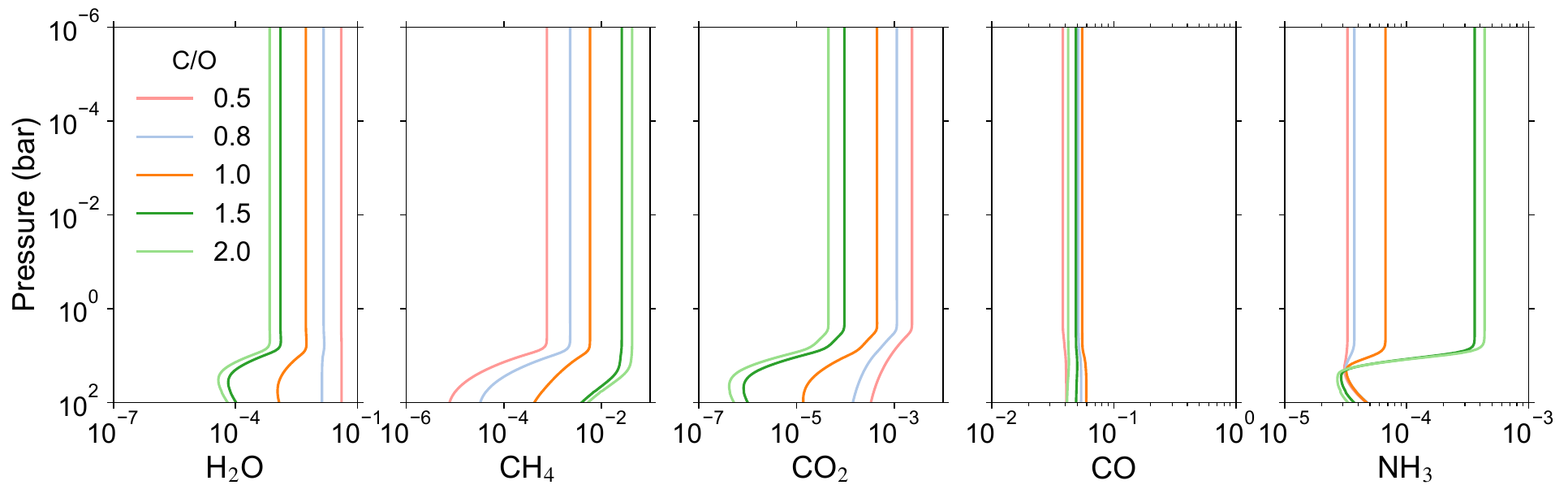}
\caption{Mixing ratio profiles of TOI-270 d atmosphere. The horizontal and vertical axes are volume mixing ratio and pressure in log scales, respectively. Only five of the eight C/O ratio scenarios are shown for clarity. The mixing ratios of reducing species (CH$_4$ and NH$_3$) monotonically increase as C/O ratio increases, while the mixing ratios of oxidizing species (H$_2$O and CO$_2$) monotonically decrease. CO concentration is roughly constant across all C/O ratios.}
\label{fig:toi270d_chem_profile}
\end{figure*}

\begin{figure*}[ht!]
\centering
\includegraphics[width=0.8\textwidth]{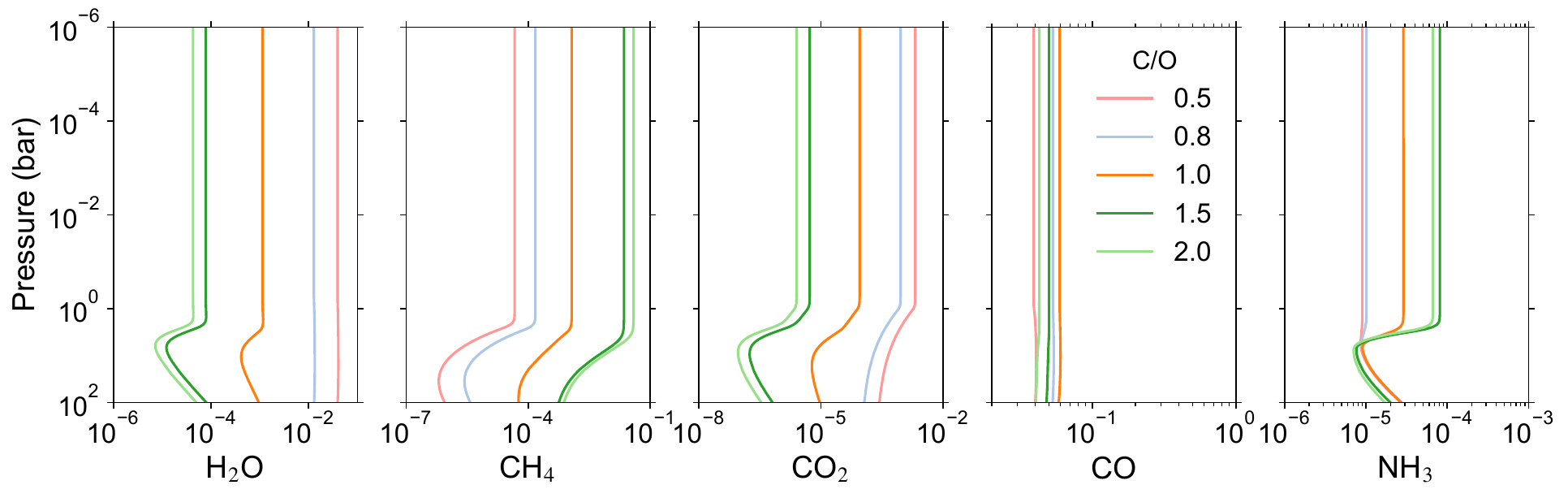}
\caption{The same as Figure \ref{fig:toi270d_chem_profile}, but for GJ 1214 b.}
\label{fig:gj1214b_chem_profile}
\end{figure*}

\begin{figure*}[ht!]
\centering
\includegraphics[width=0.8\textwidth]{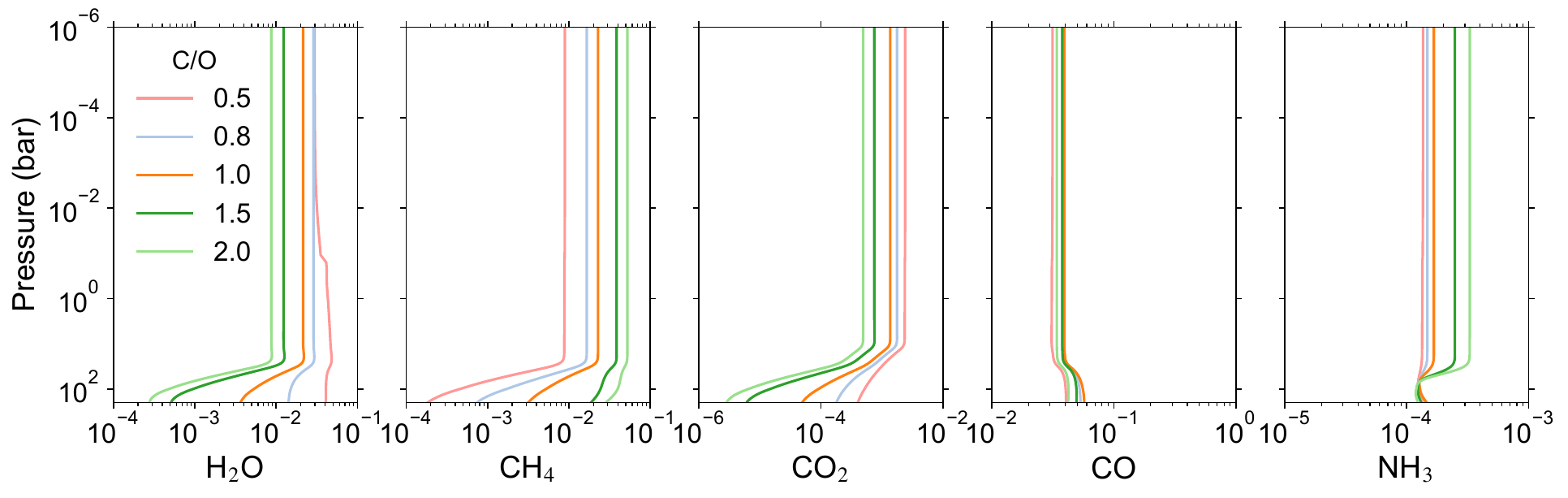}
\caption{The same as Figure \ref{fig:toi270d_chem_profile}, but for K2-18 b.}
\label{fig:k218b_chem_profile}
\end{figure*}

Here, we present the mixing ratio profiles of five spectrally important species (H$_2$O, CH$_4$, CO$_2$, CO, and NH$_3$) for TOI-270 d (Figure \ref{fig:toi270d_chem_profile}), GJ 1214 b (Figure \ref{fig:gj1214b_chem_profile}), and K2-18 b (Figure \ref{fig:k218b_chem_profile}).

\section{Simulated Transmission Spectra of GJ 1214 b} \label{sec:gj1214b_spectra}
\begin{figure*}[ht!]
\centering
\includegraphics[width=0.8\textwidth]{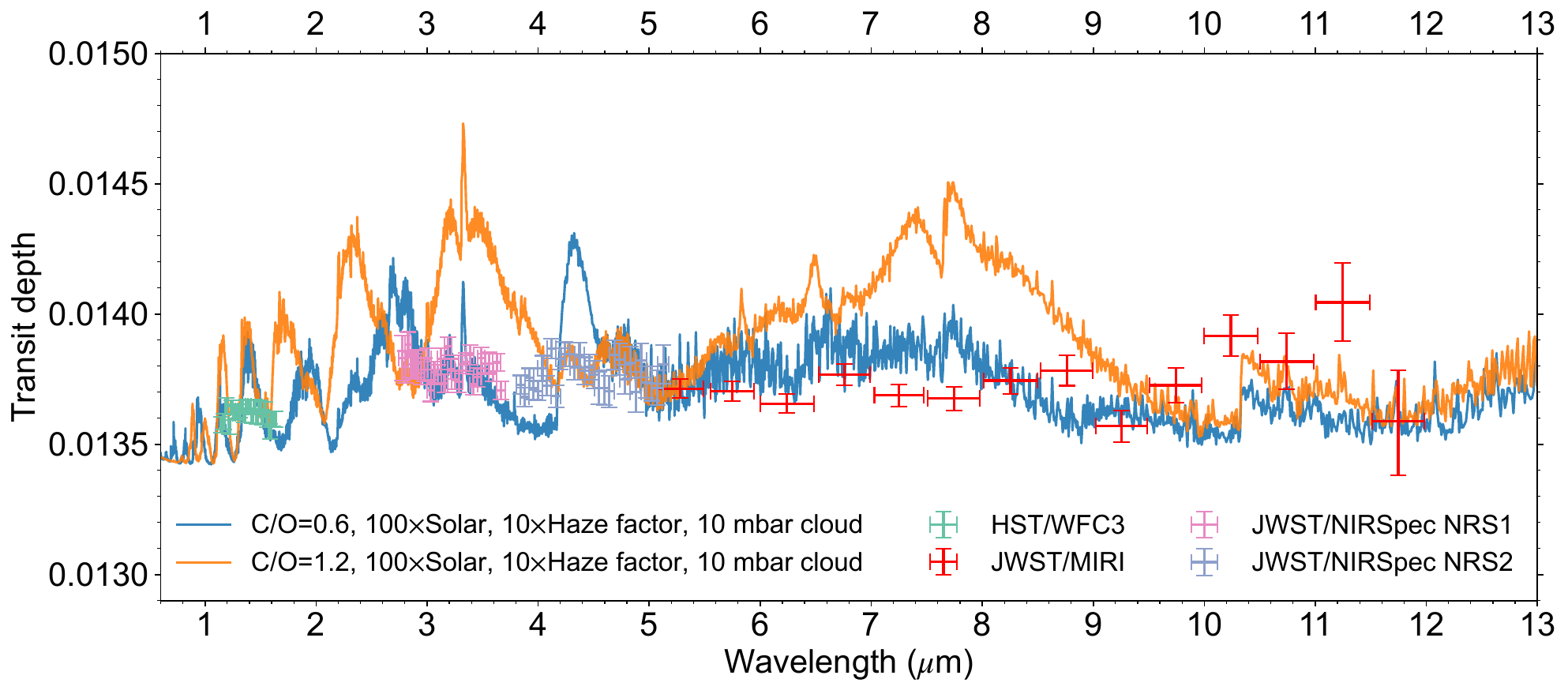}
\caption{Simulated transmission spectra of sub-Neptune GJ 1214 b compared with HST and JWST observations.}
\label{fig:gj1214b_spectra}
\end{figure*}

Here, we present the simulated transmission spectra of GJ 1214 b compared to HST and JWST observations (Figure \ref{fig:gj1214b_spectra}). The $100\times$ solar cases are visually inconsistent with the observations, which are much flatter and suggest a much higher metallicity, even if we incorporate the effects of hazes and clouds. The $1\times$ solar simulated spectra have greater scale heights than shown and are therefore even more inconsistent with observations. The $1000\times$ solar cases, while potentially compatible with the observations, are ruled out by our interior models.

\bibliography{carbon_planet}{}
\bibliographystyle{aasjournal}


\end{document}